\documentclass[fleqn,10pt]{wlscirep}
\usepackage[utf8]{inputenc}
\usepackage[T1]{fontenc}
\usepackage{siunitx} 
\usepackage{placeins}
\usepackage[version=4]{mhchem}
\usepackage{amssymb}
\usepackage{url}
\title{X-ray multibeam ptychography at up to 20 keV: nano-lithography enhances X-ray nano-imaging}

\begin{document}
\author[1]{Tang Li}
\author[2]{Maik Kahnt}
\author[3,+]{Thomas L. Sheppard}
\author[4]{Runqing Yang}
\author[1]{Ken Vidar Falch}
\author[5]{Roman Zvagelsky}
\author[4]{Pablo Villanueva-Perez}
\author[5]{Martin Wegener}
\author[1,*]{Mikhail Lyubomirskiy}

\affil[1]{Centre for X-ray and Nano Science CXNS, Deutsches Elektronen-Synchrotron DESY, Notkestr.\ 85, 22607 Hamburg, Germany}
\affil[2]{MAX IV Laboratory, Lund University, Box 118, 221 00, Lund, Sweden}
\affil[3]{Karlsruhe Institute of Technology, Institute for Chemical Technology and Polymer Chemistry, Engesserstr. 20, 76131 Karlsruhe}
\affil[4]{Division of Synchrotron Radiation Research and NanoLund, Department of Physics, Lund University, Lund, 22100, Sweden}
\affil[5]{Karlsruher Institut für Technologie, Institut für Angewandte Physik, Wolfgang-Gaede-Straße 1, D-76131, Karlsruhe, Germany}
\affil[+]{Current address: Leipzig University, Institute of Chemical Technology, Linn\'estr. 3, 04103 Leipzig, Germany}
\affil[*]{mikhail.lyubomirskiy(at)desy.de}


\begin{abstract} 
Non-destructive nano-imaging of the internal structure of solid matter is only feasible using hard X-rays due to their high penetration. The highest resolution images are achieved at synchrotron radiation sources (SRF), offering superior spectral brightness and enabling methods such as X-ray ptychography delivering single-digit nm resolution. However the resolution or field of view is ultimately constrained by the available coherent flux. To address this, the beam's incoherent fraction can be exploited using multiple parallel beams in an approach known as X-ray multibeam ptychography (MBP). This expands the domain of X-ray ptychography to larger samples or more rapid measurements. Both qualities favor the study of  complex composite or functional samples, such as catalysts, energy materials, or electronic devices. The challenges of performing ptychography at high energy and with many parallel beams must be overcome to extract the full advantages for extended samples while minimizing beam attenuation. Here, we report the application of MBP with up to 12 beams and at photon energies of 13 and 20 keV. We demonstrate performance for various samples: a Siemens star test pattern, a porous Ni/\ce{Al2O3} catalyst, a microchip, and gold nano-crystal clusters, exceeding the measurement limits of conventional hard X-ray ptychography without compromising image quality.

\end{abstract}

\flushbottom
\maketitle


\section*{Introduction}
Microscopy is a core driving force in science and technology. It plays a pivotal role in understanding the structure and function of materials by delivering access to micro- and nanoscale information. To date, sub-\SI{}{\micro\meter} resolution is routinely achieved with visible light microscopy. In contrast using electron microscopy, it is possible to exceed single-digit \SI{}{\nano\meter} resolution. Unfortunately, these methods are either constrained to retrieve only surface information (e.g. scanning electron microscopy (SEM)) of large samples, involving invasive subsampling of \SI{}{\nano\meter}-thin samples (e.g. transmission electron microscope (TEM)), or \SI{}{\nano\meter}-sized volumes for TEM tomography, or total sample destruction (e.g. FIB-SEM). Compared to the above methods, the high penetration power of hard X-rays enables non-destructive studies of relatively large samples. Among other benefits, this enables more representative investigation of complex or hierarchically structured samples while minimizing invasive subsampling. This is particularly relevant in the study of functional materials such as catalysts\cite{Cat_vis}, energy materials \cite{das_review, zenyuk_review}, or nanostructured devices such as microchips for example. Such samples are typically complex composite structures in which small sub-volumes may not be representative of the parent object.
In this context, X-ray microscopy (XRM) and computed tomography (CT) at third generation synchrotron radiation facilities (SRFs) have experienced explosive development in the last three decades.

Currently, XRM is an indispensable and flexible tool for studying extended samples (e.g. nm to cm scale) with wide resolution ranges reaching up to single-digit nm. The highest resolution measurements have been enabled by the development of lens-less imaging methods such as coherent diffraction imaging (CDI)\cite{Miao1999} and its combination with scanning probe microscopy, known as ptychography\cite{Chapman2010,Dierolf2010, Pfeiffer2018, grote2022, Taphorn2022}. In hard X-ray ptychography, the sample is irradiated by a confined coherent beam, with the diffraction signal recorded by a detector. By measuring the beam position on the sample and recording the diffraction signal, the sample complex transmission function and the probing beam are iteratively reconstructed. As a lens-less imaging technique, ptychography exploits focusing optics purely to increase photon fluence on the sample and speed up data collection. Another aspect of ptychography is that the incoming beam required to be fully or nearly fully coherent and typical coherent fraction of the beam at modern SRFs of 3rd generation is less than \SI{1}{\percent}, this requires wasting nearly \SI{99}{\percent} of the incoming beam. Notably, ptychographic measurements are also possible with electron microscopy where they operate at even higher resolution than conventional TEM, although still limited to very small sample volumes \cite{Cryo_electron_ptycho2023,Ding2022}.

The major shortcoming of ptychography is that resolution is ultimately limited by the X-ray beam flux at the sample. Consequently, assuming a certain flux and finite beamtime allocation at a given SRF, either the sample size or target resolution must be constrained. This hinders the core advantage of XRM (i.e. high-resolution imaging of extended samples) in either of two ways. Firstly, by limiting the sample size in order to limit measurement duration. This can force invasive subsampling and potentially unrepresentative imaging of complex samples. Secondly, by compromizing the measurement resolution in order to increase sample size, therefore rendering many \textit{in situ} studies of transient processes as unfeasible due to lack of sensitivity to small changes in the sample. While beamtime availability therefore restricts measurement of extended samples at maximum resolution, an additional challenge comes from X-ray attenuation. As sample thickness increases, higher photon energies are needed to penetrate thicker samples. High energy ptychography is therefore an attractive prospect, due to minimal attenuation and coincidentally decreased beam damage due to smaller absorption cross-sections. However, considering that the available coherent fraction of photons at SRFs is inversely proportional to energy, high-energy ptychography is challenging in practise due to insufficient flux. For this reason, few studies report on X-ray ptychography at energies above 17 keV\cite{daSilva:gb5085}.

Resolving the issues above would enable high-resolution imaging of extended samples on feasible timescales. However, performing ptychography under these conditions constitutes one of the most challenging issues in modern X-ray imaging. One approach to resolve this is with fourth-generation diffraction-limited SRFs offering increased coherent flux in the hard X-ray regime. However, such facilities are prohibitively expensive and time-consuming to construct. The central issue of limited beamtime availability cannot feasibly be resolved by simply building more SRFs. An alternative approach developed in recent years involves the inclusion of previously unusable photons into ptychographic experiments and reconstruction algorithms. This approach is called multibeam ptychography (MBP), and is based on dividing an incident X-ray beam into multiple beams which are in themselves coherent, but mutually incoherent. MBP was first demonstrated in 2017 with visible light\cite{bevis2017}, and later with X-rays\cite{Yao2020, Hirose2020, Wittwer21, Lyubomirskiy22}. Since multiple regions of the sample are imaged simultaneously, this speeds up ptychography with increasing number of beams on a linear scale. This is equivalent to performing multiple conventional ptychography measurements simultaneously. Development of MBP is therefore of great interest in the context of high-resolution X-ray imaging on realistic samples or with rapid acquisition rates such as in functional materials, catalysis, or nanofabrication.

The major challenge in X-ray MBP involves using a larger fraction of the incoming beam at higher energies where gains in measurement speed and sample size are the most relevant. Various types of X-ray optics have been proposed for this purpose, such as array of Fresnel zone plates (FZPs)\cite{Yao2020}, focusing mirrors with slits\cite{Hirose2020}, and compound refractive lenses (CRLs)\cite{Wittwer21, Lyubomirskiy22}. The criteria for efficient MBP with hard X-rays are: (i) ptychography requirements need to be met, e.g. confined and coherent illumination, (ii) optics need to be highly efficient. Since no off-the-shelf optics exist for MBP, custom solutions are needed. While FZP have well-established manufacturing processes, they are unsuitable due to their low efficiency (below \SI{10}{\percent}) at X-ray energies above 12 keV\cite{Gorelick2011-rh}. 
Mirrors are only usable in combination with other optics (e.g. FZP, slits) due to mechanical constraints, which ultimately hinders the photon efficiency and practicability of such schemes. The proven most feasible option is double concave CRLs \cite{Lengeler2002, Schroer2003, Petrov17, Lyubomirskiy2019}. Another challenge in MBP is that the beams must be sufficiently unique in phase and/or amplitude in order to achieve robust separation of overlapping signals from multiple beams at the detector. It has been shown that specifically designed different phase plates\cite{Seiboth19} added to each lens stack can achieve this\cite{Lyubomirskiy22}.

The pivoting point in the development track of MBP was the application of enabling technology -- 3D laser two-photon absorption printing technique for manufacturing focusing optic arrays was used to create focusing elements out of polymer with full geometric freedom and high precision\cite{Wegener22}. Then, the highest known photon utilization for MBP of \SI{98}{\percent} was achieved  at energies of 7 and 9 keV with doubly concave CRLs in tightly packed tower arrays\cite{Lyubomirskiy22, Lyubomirskiy2019}. A radius of curvature of a single parabolic surface of single-digit \SI{}{\micro\meter} was achieved. To perform MBP at 20 keV (compared to previous measurements at 7 keV), the focusing power of a single lens tower requires an order of magnitude smaller effective lens curvature (effective curvature = single lens curvature/number of lenses), from state-of-the-art \SI{0.83}{\micro\meter}\cite{Lyubomirskiy22} down to \SI{100}{\nano\meter} level as the lens focusing power decreases quadratically with the X-ray photon energy. Otherwise the focused beam size or "pupil" at the detector will be reduced which can lead to earlier pixel saturation at constant flux rates.
The low optical density of polymers is therefore a major weakness of laser printed optics in the hard X-ray regime. On the other hand, no alternative manufacturing process can deliver tightly packed arrays of refractive lenses. To achieve large focusing power more individual lens elements have to be stacked together, leading to an increased lens tower aspect ratio and consequently more strict requirements for manufacturing precision. This sets very challenging constraints on manufacturing as the lens tower aspect ratio reaches 100:1, keeping in mind that manufacturing precision is crucial to the overall success of the MBP measurements.

Unlike previous experiments with MPB, most challenging of which were constrained to a maximum of 3 parallel beams\cite{Lyubomirskiy22} and softer X-rays, here we report on the first application of MBP with up to 12 beams at irradiation energies of 13 and 20 keV. This was accomplished by developing a lens array using cutting-edge laser printing technology. The quality and robustness of MBP are demonstrated on a range of sample systems, including traditional test patterns, a microchip, porous catalyst structures, and gold nano-crystal particles. Hard X-ray MBP is therefore demonstrated as a method of the highest potential to achieve extended imaging of large samples at the nanoscale, which is temporally unfeasible with state-of-the-art single-beam X-ray ptychography.


\section*{Results}

\subsection*{Lens arrays design, manufacturing and corrections}
\begin{figure}[ht]
\centering
\includegraphics[width=0.6\linewidth]{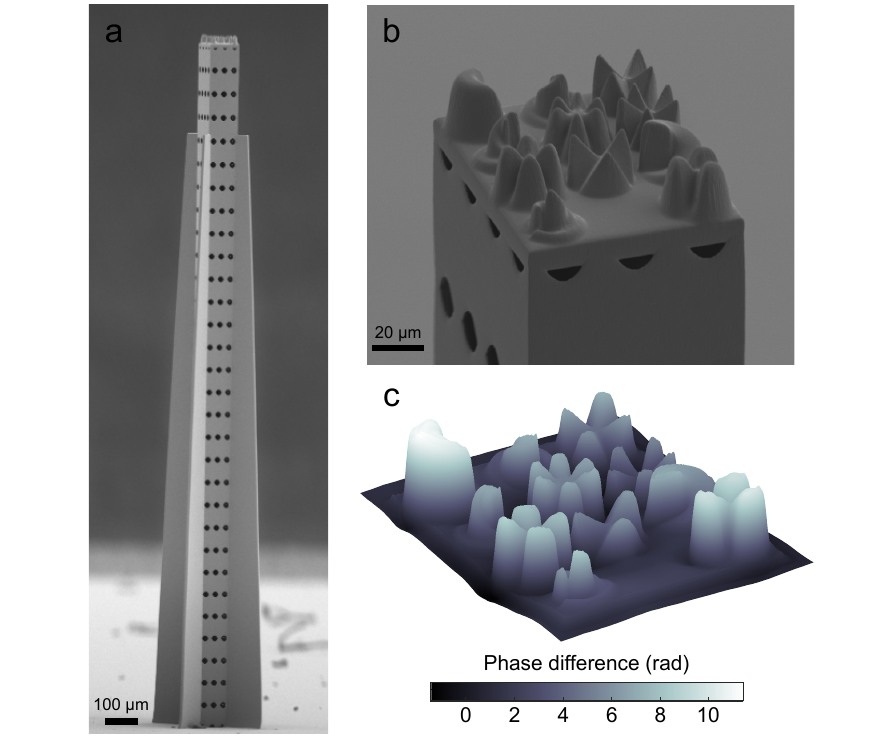}
\caption{(a) SEM image of a lens array of 30-fold lens towers in $3\times4$ grid for 13 keV photon energy with phase plate on top. (b) SEM close-up of the same phase plate. (c) \textit{In situ} QPI of a separately printed phase plate showing phase difference between unpolymerized and polymerized photoresist. The phase measured with a wavelength of \SI{630}{\nano\meter}.}
\label{fig:sem}
\end{figure}

Two lens arrays were used in the current work, consisting of 12 individual 30-fold lens towers in a $3\times4$ grid ($V\times H$) (Fig. \ref{fig:sem}) for experiments at 13 keV, and corresponding 40-fold lens towers for experiments at 20 keV. The lens arrays were manufactured and corrected, with experimental validation in between. For this purpose, optics performance was assessed with 13 keV irradiation at the synchrotron. Each lens tower in the array was individually characterized with single beam ptychography, and the reconstructed complex wave-fields were propagated numerically. After the tests, it was discovered that in the first iteration, the two central towers (from $3\times4$ grid) had different focal lengths with respect to the border towers, possibly due to the dose accumulation during laser printing. 
This led to mismatch of the size of the two central beams and the remaining beams and directly affected the efficiency of MBP scans, since the scan step size is reduced and measurement time therefore increased. Due to sampling requirements overlap of the irradiated sample regions in adjacent scan positions should not exceed \SI{60}{\percent}. The experimental tests of the corrected lens tower arrays (second iteration) at 20 keV showed expected joint lens towers performance, providing sufficient experimental conditions for executing MBP with different beam arrangements.

Fig. \ref{fig:sem}(a) shows an example of a printed lens tower array for 13 keV photon energy. A close-up of the phase plate on top of the lens system is shown in Fig. \ref{fig:sem}(b). The quality of integrated phase plates on top of each lens tower is especially important since the separation of the signals from different beams relies on differences in their phase and/or amplitude. For this purpose, we characterized the phase plate using a recently developed \textit{in-situ} quantitative phase imaging (QPI) technique \cite{qpi}. A resulting \textit{in-situ} phase topography map measured on a separately printed phase plate is depicted in Fig. \ref{fig:sem}(c).

\FloatBarrier


\subsection*{Multibeam ptychography results}
\begin{figure}[ht]
\centering
\includegraphics[width=0.95\linewidth]{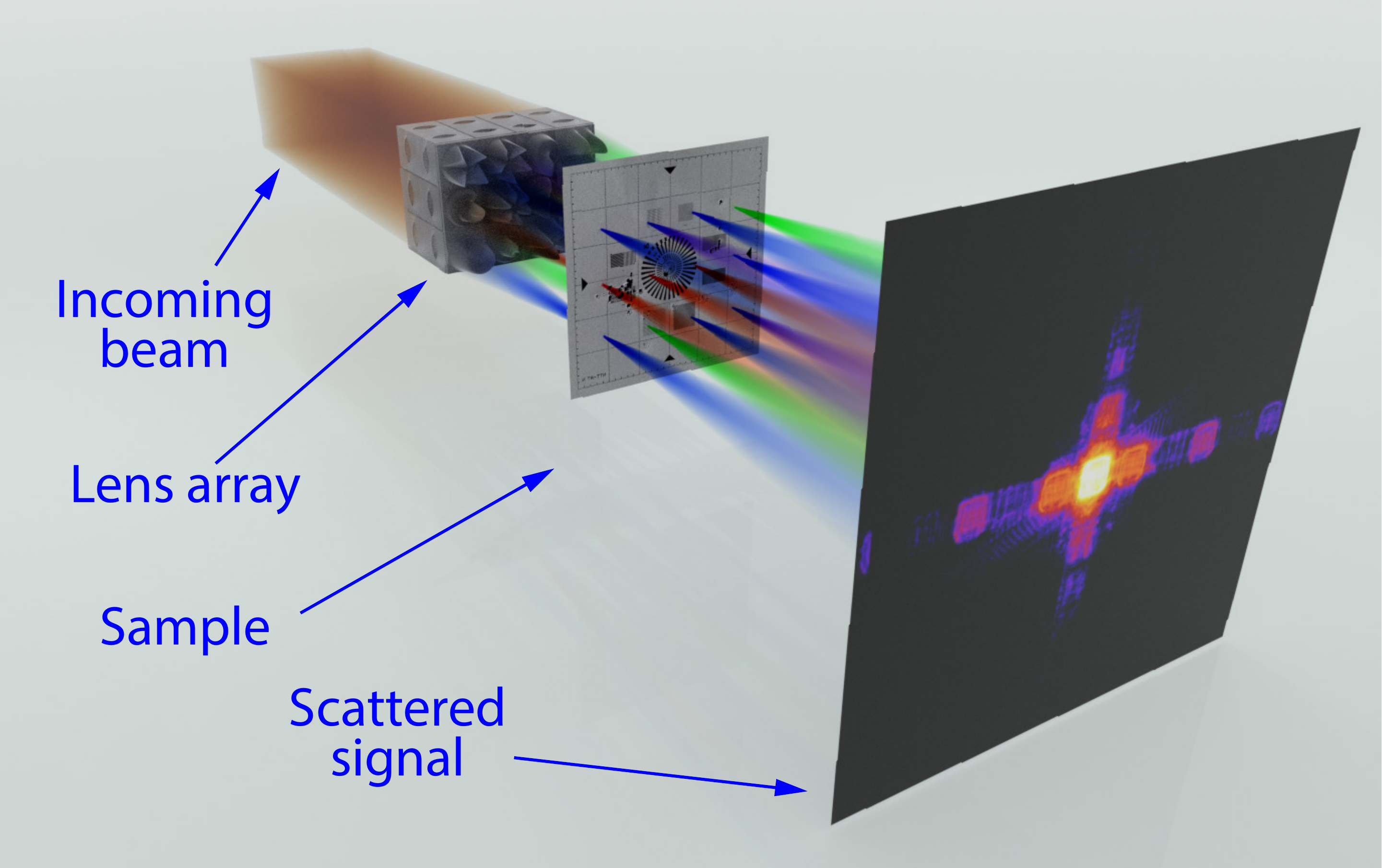}
\caption{Scheme of the performed multibeam ptychography showing separation of primary beam into individual coded beams which simultaneously irradiate multiple sample points. Scattered beams are propagated to the detector analogously to conventional hard X-ray ptychography.}
\label{fig:scheme}
\end{figure}

The general measurement scheme for MBP is depicted in Fig. \ref{fig:scheme}. Experiments were performed at two beamlines: P06 at PETRA III (13 keV) and ID13 at ESRF-EBS (20 keV). A similar lens array design (see above) with 30-fold and 40-fold lens towers were used, respectively. The total desired number of illuminated lens towers producing beams was selected using slits upstream of the lens array (not shown in the scheme). For all beams combinations the scan range was \SI{35}{\micro\meter}\texttimes  \SI{35}{\micro\meter} to cover the distance between neighboring lens towers of \SI{30}{\micro\meter}  with \SI{5}{\micro\meter} of overlap of sample regions scanned by adjacent beams. 

\subsubsection*{Multibeam ptychography at 13 keV: 6 and 12 beams}
\begin{figure}[ht]
\centering
\includegraphics[width=0.85\linewidth]{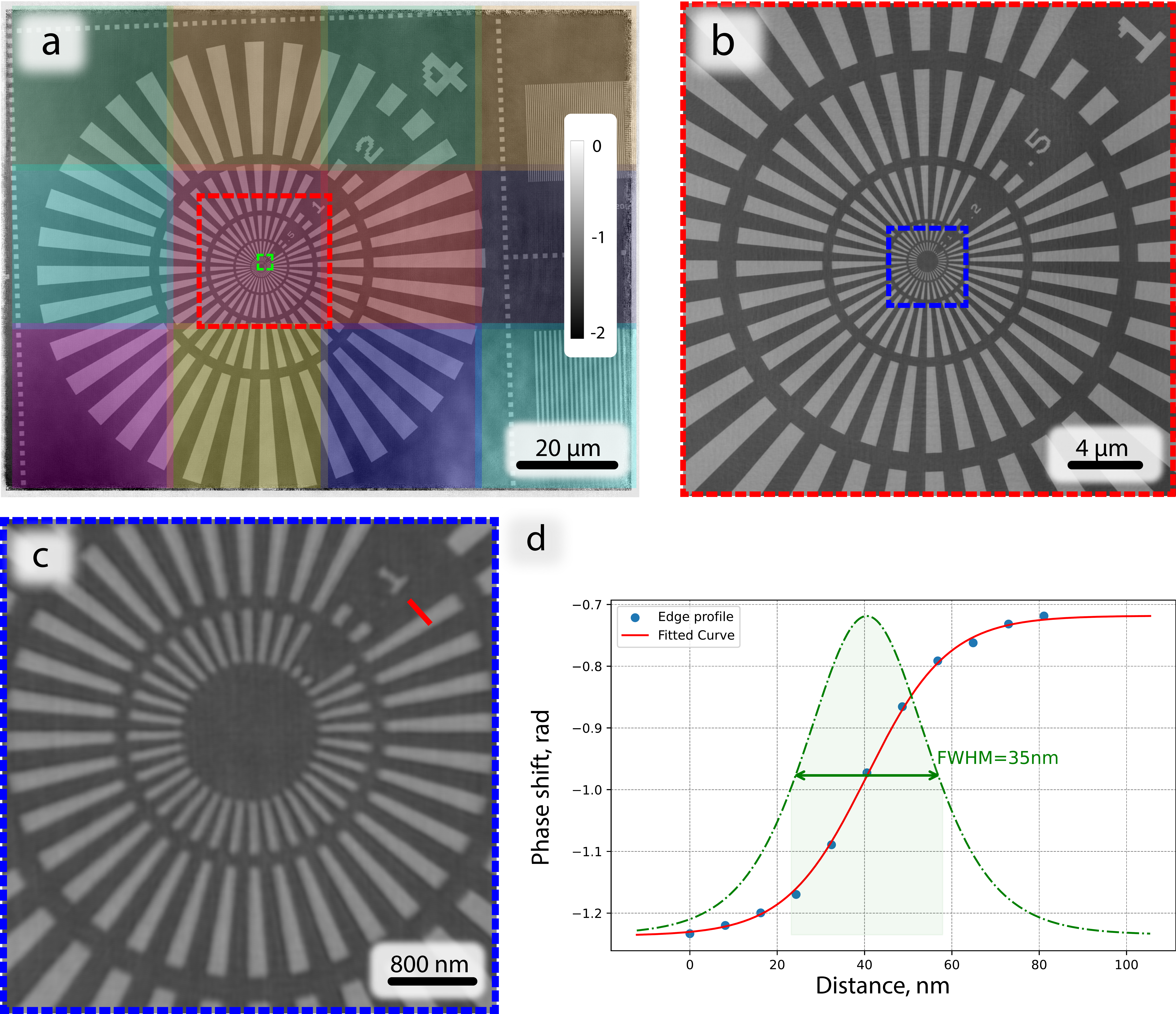}
\caption{Reconstructed object phase from the 12 beam measurement of the Siemens star test pattern at \SI{13}{\kilo\eV}; (a) overall image, colored squares represent regions irradiated by each individual beam respectively; (b) magnified central region partially scanned by two neighboring probes; (c) magnified region with smallest features - \SI{50}{\nano\meter}, red line indicates the position for the line profile; (d) line profile of the edge estimating resolution; The color bar represents phase shift in radians. The pixel size in reconstruction is \SI{8}{\nano\meter}.}
\label{fig:siemensstar}
\end{figure}

A series of samples with different structural features and/or complexity were used to validate the performance of the MBP measurements and MBP reconstruction algorithms. Firstly, a Siemens star XRESO-50HC\cite{XRESO} manufactured by NTT-AT with smallest features of \SI{50}{\nano\meter} was imaged. For the 12 beams arrangement a full lens array was illuminated. This resulted in an effective scanned area of \SI{95}{\micro\meter}\texttimes  \SI{125}{\micro\meter}. The reconstructed phase image of the Siemens star is depicted in Fig. \ref{fig:siemensstar}, with color shading used to indicate the 12 separate regions of the sample irradiated by each of the 12 probes respectively. This color indication made it easy to inspect regions where signals from different probes needed to be "stitched" by the reconstruction algorithm. This information was used to check for possible artefacts or errors in the reconstruction process. For example, the magnified central area was scanned with two individual probes (Fig. \ref{fig:siemensstar}(b)), whereby no visible artifacts are present in the reconstruction, and the \SI{50}{\nano\meter} bars are clearly resolved (\ref{fig:siemensstar}(c)). This indicates that the diffraction signals from different sample parts were robustly deconvoluted by the reconstruction algorithm, providing high-quality reconstructions.
 Moreover, the line edge profile (Fig. \ref{fig:siemensstar} (d)) indicates a resolution of \SI{35}{\nano\meter} FWHM, which is comparable with the result of a single beam reconstruction of \SI{34}{\nano\meter} from the scan taken for the characterisation of single probes with normalized statistics per irradiating beam -- see comparison in supplementary materials.

In summary, MBP is therefore comparable in terms of resolution as conventional single beam ptychography\cite{Lyubomirskiy22}, while facilitating rapid scans over large fields of view through simultaneous measurement of multiple sample positions.

\begin{figure}[ht]
\centering
\includegraphics[width=0.95\linewidth]{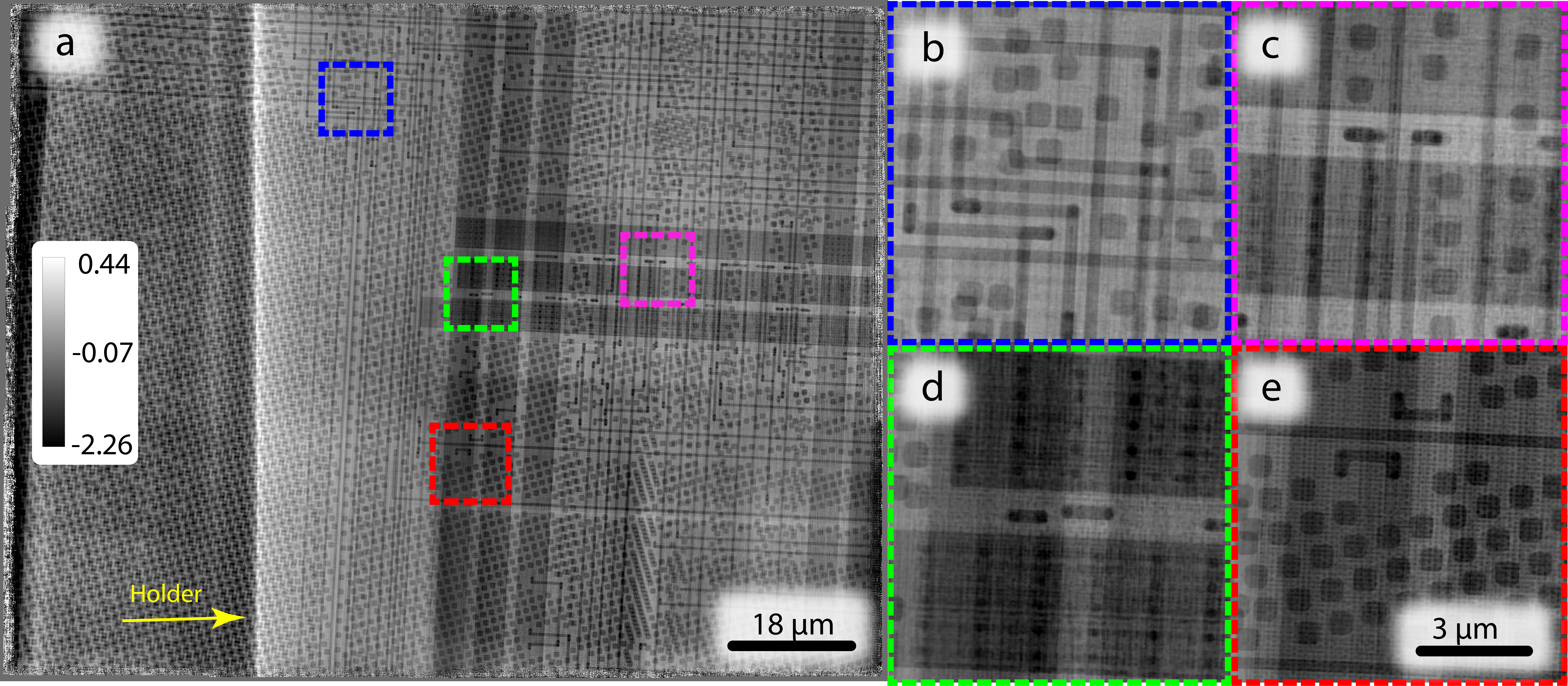}
\caption{Reconstructed object phase from the 12 beam measurement of the microchip at \SI{13}{\kilo\eV}; (a) overall image; (b)-(e) magnified regions indicating small features; The color bar represents phase shift in radians. The pixel size in reconstruction is \SI{16}{\nano\meter}.}
\label{fig:microchip}
\end{figure}

To further assess the capabilities of MBP, samples with more diverse features were examined. These represent real objects which may form the basis of potential application areas for MBP. Firstly, a microchip containing heterogeneous pattern structures of circuits, transistors, and contacts spread over several layers was scanned with 12 beams in the same manner as described above. As an extended planar object, the microchip represents a potential application which is ideally performed by scanning of an array of beams in MBP instead of just one beam in conventional single beam ptychography. The reconstructed microchip is depicted in Fig. \ref{fig:microchip}. The sample visibly contains a large number of different scale features, from several \SI{}{\micro\meter} -- a "trench" originating from the top of the image (Fig. \ref{fig:microchip}) (a)) to tens of \SI{}{\micro\meter} dots Fig. \ref{fig:microchip}) (e)). All of these features are clearly visible in the reconstructed data with no noticeable artifacts. On the magnified images (Fig. \ref{fig:microchip}) (b)-(d)) it is possible to see the smallest features with more details. Another noticeable large scale feature is long range phase shift variation over the whole sample with a horizontal gradient, this is the structure of the sample holder that the sample was fixed on and that was in the beam path.

\begin{figure}
    \centering
    \includegraphics[width=0.8\linewidth]{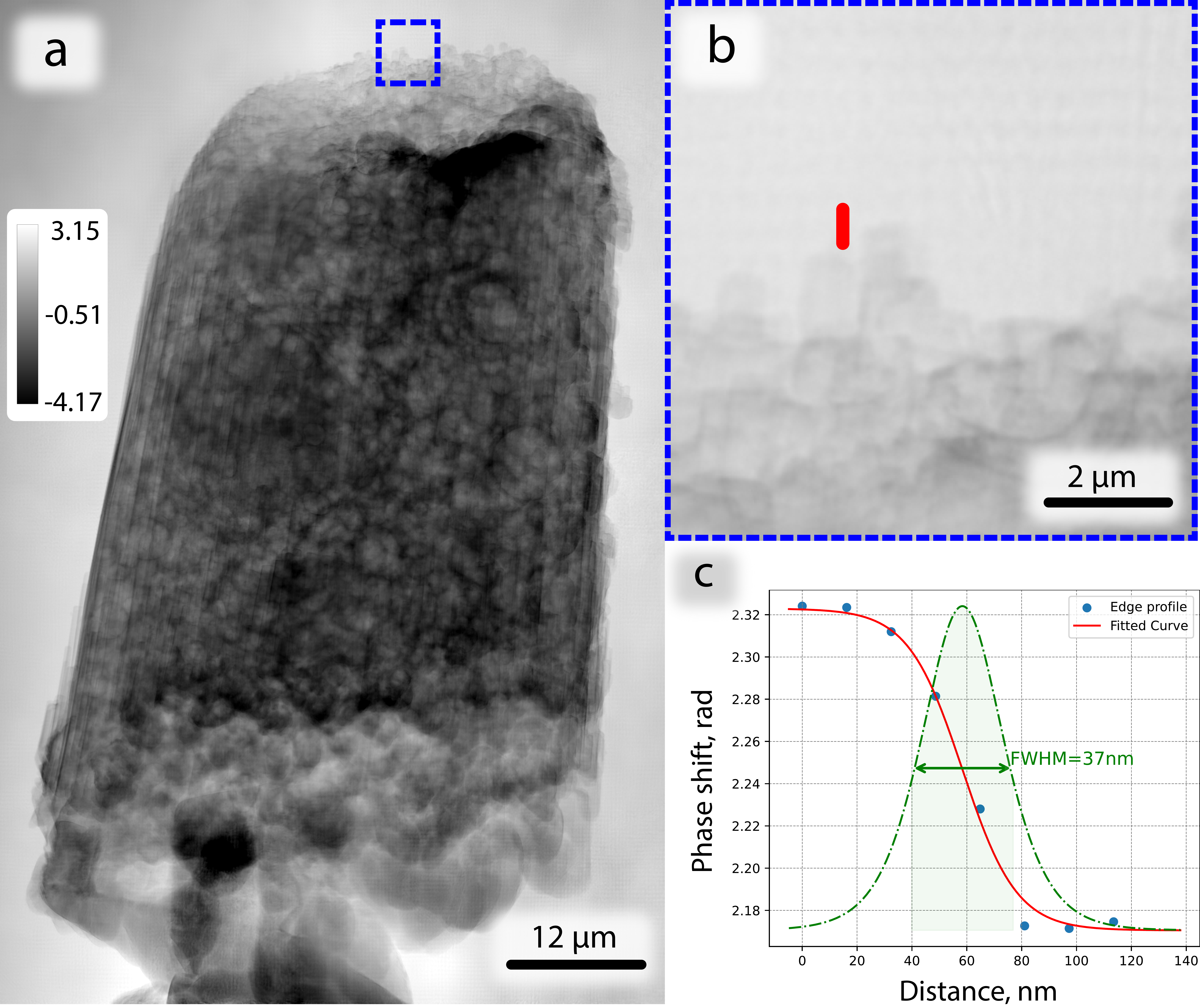}
    \caption{Unwrapped reconstructed object phase from the 6 beam MBP measurement on the Ni/\ce{Al2O3} catalyst sample at \SI{13}{\kilo\eV}; (a) overall image; (b) magnified region; (c) edge line profile; The color bars represent phase shift in radians. The pixel size in reconstruction is \SI{16}{\nano\meter}.}
    \label{fig:tomo}
\end{figure}

A second potential application area of MBP is in the study of complex composite samples with less organized or more random structural features. These may be represented by energy materials (e.g. batteries, fuel cells) or solid catalyst samples for industrial chemistry applications. These are often composite materials, while entire samples can even exceed mm-cm scale. In this context, MBP is particularly interesting as it enables larger fields of view which may be more representative of the parent sample. This is in principle performed without compromising on resolution. Here, a sample of hierarchically-porous Ni/\ce{Al2O3} catalyst was prepared as a cylinder of approx. \SI{45}{\micro\meter}. Preparation was performed by focused ion beam (FIB) and the sample was placed on a tomography pin, as described in previous work\cite{weber_tomo}. The sample was designed for 6 beam arrangement $2 \times 3$ ($H\times V$) with corresponding FOV of \SI{65}{\micro\meter}\texttimes  \SI{95}{\micro\meter}. A single projection showing the reconstructed phase image is depicted in Fig. \ref{fig:tomo} (a). The total phase shift in the object exceeded 2$\pi$, and thus the reconstructed image initially contained phase wraps. These were unwrapped during post-processing. The final reconstruction clearly indicates the presence of the expected complex interior pore network. This was thoroughly characterized in previous ptychographic X-ray computed tomography (PXCT) studies and consists of a combination of mesopore (2-20 \SI{}{\nano\meter} diameter, not resolved) and macropore (>\SI{50}{\nano\meter} diameter, resolved) features, with the latter extending up to \SI{2.5}{\micro\meter}\cite{weber_tomo}.
Based on the line edge profile in Fig. \ref{fig:tomo}(b-c) the achieved spatial resolution was estimated to be \SI{37}{\nano\meter}. These results are directly comparable to previous PXCT measurements, in which a \SI{15}{\micro\meter} diameter sample with broadly similar structural features was measured with around \SI{56}{\nano\meter} 3D resolution\cite{weber_tomo}. 
These results indicate that MBP may in principle be extended to long-range tomographic measurements which greatly exceed the feasible FOV of conventional single beam ptychography due to time constraints.

\FloatBarrier

\subsection*{Multibeam ptychography at 20 keV: 6 beams}
To examine the possibility to perform MBP at even higher energies, additional experiments with 20 keV irradiation were performed at the ESRF-EBS beamline ID13. It should be noted that despite the high spectral brightness of the upgraded source, the experiment suffered from low photon statistics; at that energy only the Maxipix GaAs detector without integration mode had sufficient quantum efficiency, limiting the total intensity per pixel to 12518 counts. Furthermore, the control software was not able to provide multiple exposures per point scanning regime. Consequently, the recorded intensity per beam during MBP experiments could not reach the same level as with a single beam, which negatively affected the achieved resolution due to the lower signal-to-noise ratio (SNR). Because of this, the number of beams was limited to 6, which showed sufficient quality in the reconstructed object. The supporting information details how the resolution of the reconstructed object degraded with an increased number of beams used. The following examples therefore indicate the successful application of high-energy MBP, but the quality of the results under-represent the potential performance of MBP in future high-energy experiments with a more appropriate detector such as CITIUS\cite{CITIUS} and JUNGFRAU\cite{JUNGFRAU}.

\begin{figure}
    \centering
    \includegraphics[width=0.6\linewidth]{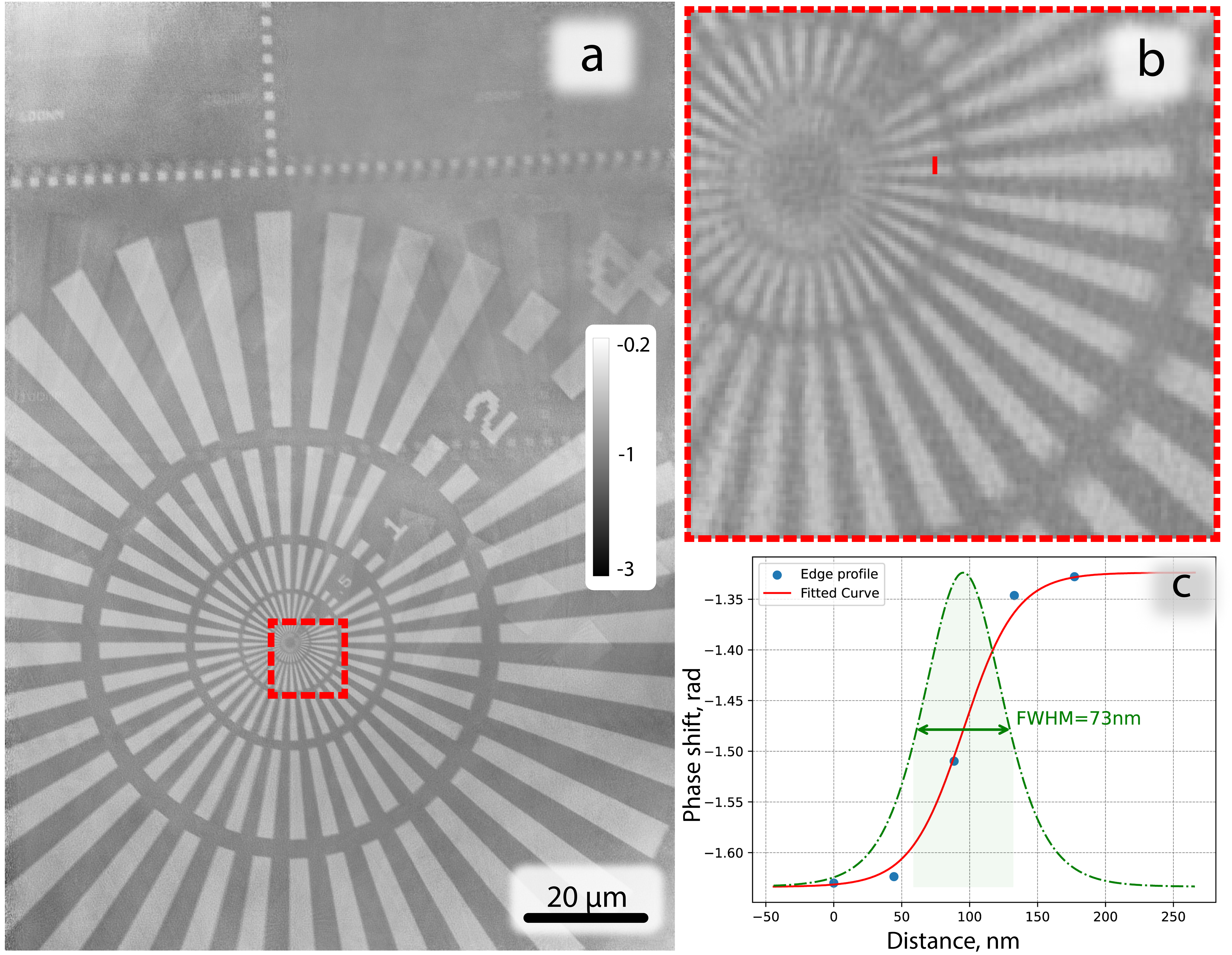}
    \caption{Reconstructed object phase from the 6 beam measurement of the Siemens star test pattern at \SI{20}{\kilo\eV}; (a) overall image; (b) magnified central region; (c) line profile. The pixel size in reconstruction is \SI{44}{\nano\meter}.}
    \label{fig:Siemens Star 6beam}
\end{figure}

A Siemens star resolution test chart sample was again taken as an initial measurement at 20 keV. The reconstructed image with 6 beams (2x3 arranged HxV) is depicted in Fig. \ref{fig:Siemens Star 6beam}. The estimated resolution according to a line profile of the edge was \SI{69}{\nano\meter}. This coincides with the fact that the smallest lines and spaces (\SI{50}{\nano\meter}) in the center of the test pattern can not be resolved, while the second smallest features of \SI{100}{\nano\meter} were clearly resolved. The image has artifacts in the top region originating from the lack of high diversity of features, leading to reduced quality of the reconstruction.

\begin{figure}
    \centering
    \includegraphics[width=0.95\linewidth]{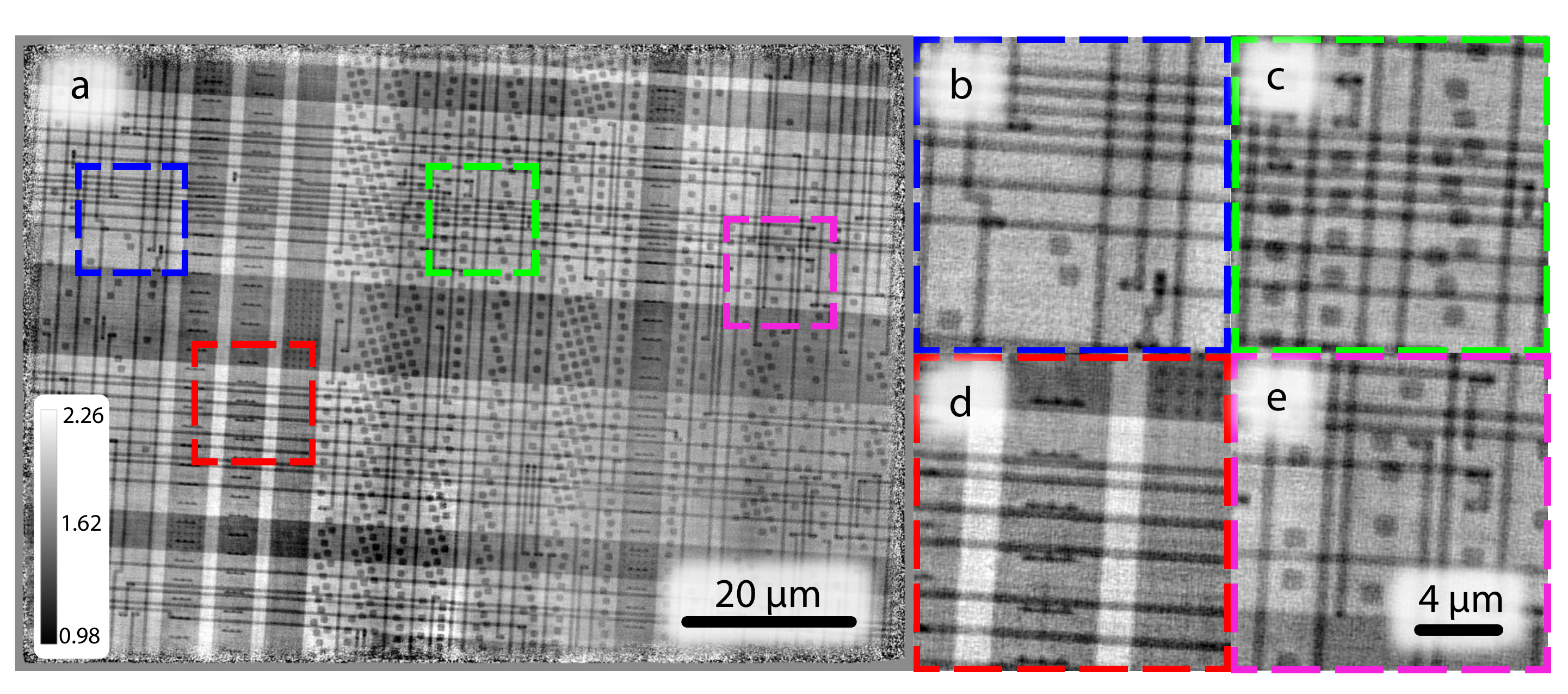}
    \caption{Reconstructed object phase from the 6 beam measurement of the microchip at \SI{20}{\kilo\eV}; (a) overall image; (b)-(e) magnified regions indicating small features; The color bar represents phase shift in radians.The pixel size in reconstruction is \SI{44}{\nano\meter}.}
    \label{fig:chip20}
\end{figure}

As with previous experiments at 13 keV, additional samples were chosen with increased structural complexity. Firstly, a microchip with regularly spaced highly diverse features was again measured. The reconstructed phase shift of the microchip sample is depicted in Fig. \ref{fig:chip20}. At 20 keV irradiation, the microchip sample has a smaller cross-section, resulting in comparably weaker scattering than at 13 keV. In combination with the reduced photon statistics discussed previously, this led to a lower resolution compared to the 13 keV experiment. Despite this, the image had no noticeable reconstruction artifacts, while different length-scale sample features were robustly reconstructed. The major difference with the 13 keV experiment is that there was no Al holder in the beam, thus there are no long range features in the background of the reconstructed object and all visible structures are a result of the internal components of the microchip.
As a final test sample, gold nano-crystal clusters were prepared on a SiN membrane with sizes ranging between \SI{1}{\micro\meter} and \SI{6}{\micro\meter} in size. These represent dispersed sparse objects which are highly scattering and are therefore are interesting test object for MBP since they are non-contiguous objects on otherwise featureless background.
In summary and despite reduced photon statistics, MBP is shown to be robust even at the high photon energy of 20 keV.

\begin{figure}
    \centering
    \includegraphics[width=0.95\linewidth]{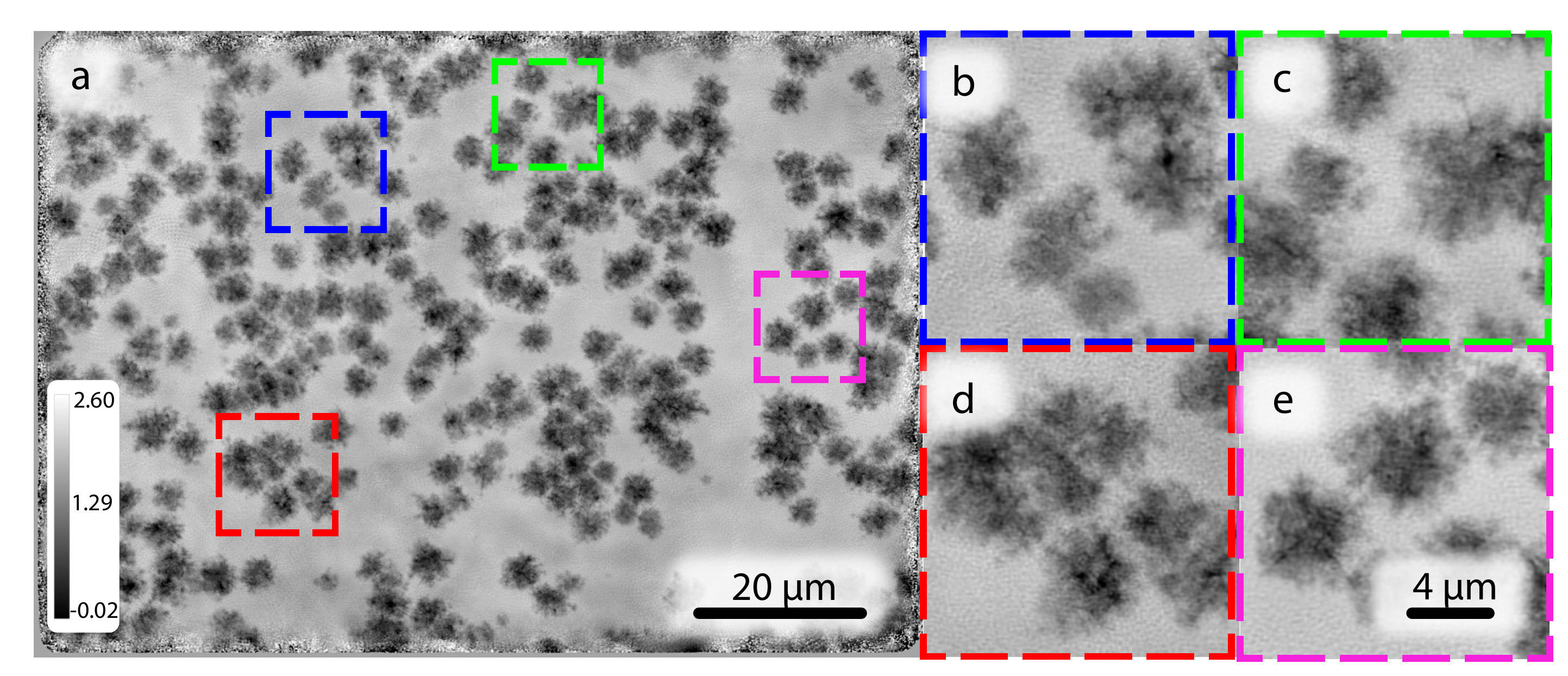}
    \caption{Reconstructed object phase from 6 beam measurement of gold nano-crystal cluster on $Si_3N_4$ membrane at \SI{20}{\kilo\eV}; (a) overall image; (b)-(e) magnified regions indicating small features; The color bar represents phase shift in radians. The pixel size in reconstruction is \SI{44}{\nano\meter}.}
    \label{fig:Au_particles}
\end{figure}

\FloatBarrier


\section*{Discussion}
Compared to contemporary imaging methods with visible light, X-rays, or electrons, hard X-ray ptychography in 2D or 3D offers significant advantages in the study of relatively large objects at the nanoscale. Despite this, a core criticism of ptychography and scanning probe methods more generally, is that they tend to be slow techniques due to the need to scan multiple positions\cite{batey2022}. By demonstrating successful application of MBP with up to 12 beams and at energies of 13 keV and 20 keV, we present a pathway to extend the use of ptychography to larger samples or more rapid acquisition rates. Crucially, the presented MBP reconstructions achieved the same spatial resolutions as conventional single beam prychography reconstructions using similar experimental parameters\cite{Lyubomirskiy22}, see supplementary information.
It should be noted that the size and overall design of the conventional test pattern -- Siemens star, which has been used for resolution comparison, is not ideally suited for assessment of the performance of MBP. This is evident from the image (Fig. \ref{fig:siemensstar}(a)), wherein the total imaged area is much larger than the high-resolution zone at the center of the Siemens star, which is usually being measured for resolution assessment. 

Electron microscopy is often regarded as a universal imaging tool in chemistry, physics, and materials science. While it can provide remarkable single-digit \SI{}{\nano\meter} resolution, or even single atoms using electron ptychography, the major drawback comes from sample size limitations. This in principle presents a fundamental physical constraint that cannot be overcome, necessitating the use of hard X-rays for measuring extended samples. In case of MBP it is now proven that it can perform non-destructive imaging of extended samples with lateral size exceeding the travel range of typical scanning stages and with resolution on the level of single beam ptychography. 

The need to measure extended samples at high-resolution is most urgent in the study of functional materials and nanodevices, where invasive subsampling may lead to unrepresentatively small volumes. In fields such as energy materials, catalysis, and microelectronics for example, the ability to measure samples of \SI{100}{\micro\meter} or greater with resolutions of \SI{20}{\nano\meter} or below will open unprecedented possibilities for the accurate characterization of nanoscale structure. In addition to simply measuring larger samples, the use of MBP may in particular enable rapid scans of smaller samples, facilitating the use of in situ methods to image transient processes with greater accuracy\cite{grote2022,weber_tomo,meirer2018}. A further application is in high throughput imaging of samples with large structural variations, such as in heterogeneous catalysis\cite{Bossers2021} in which there can be high variation between individual samples\cite{Bare2014}. All of the above represent current challenges in X-ray imaging, all of which may be overcome by MBP as demonstrated here. In particular, the extension of MBP to tomographic regimes would be particularly attractive for such samples due to their complex 3D architecture. However tomographic experiments will necessitate the use of high-energy X-rays, to minimize both beam attenuation and extensive phase artifacts such as wraps or vortices in larger samples\cite{daSilva:gb5085}. Therefore successful demonstration of MBP at 20 keV is a significant step towards future MBP tomography studies of complex solid matter.

The 3D two-photon printing technique for optics manufacturing applied here has now reached the performance level to deliver high aspect ratio structures with remarkable precision. Here optics were manufactured with an aspect ratio exceeding 100:1 (single lens tower). This is especially important for MBP at higher energies, as it allows maintaining sufficient focusing power and provides sufficient quality of phase coding plates, which is crucial for successful separation of diffraction signals from different beams. Further optics development for MBP will eventually move towards reducing single-lens aperture sizes while maintaining sufficient NA, which is dictated by the demands of the acquisition speed. The scanning time in MBP is ultimately limited by the total flux per beam, and the scan range required to cover the full distance between neighboring beams. While the synchrotron ultimately defines the total flux per beam, the scanning range can be varied by reducing the distance between neighboring beams. By reducing the lens aperture, it will be possible to vary the scanning range and thus the scanning time significantly. Additionally, prefocusing may be used to match the coherent length of the synchrotron beam and the lens aperture. In this way, it will be possible to image objects of remarkable size while performing very small scans. Already now, the imaged area of several objects described here exceeded the maximum range of the used scanning stage (\SI{100}{\micro\meter}). The ability of the 3D two-photon printing technique to quickly create virtually any design optics constitutes it as the flexible, precise, and reliable tool for manufacturing on-demand tailored optics for specific needs of particular experiment and SRF.

Currently, a major direction in MBP development is performing it at even higher energies >25 keV. Here the potential speed gains are especially high, considering the presence of fourth-generation synchrotrons such as MAX IV\cite{MAXIV} or ESRF-EBS, along with future projects such as PETRA IV\cite{PETRAIV}, SPRING 8-II or Diamond II. At such SRFs, the beam coherence fraction at higher energies (>25 keV) will be comparable with what current sources can achieve at <10 keV. This will open the avenue for imaging of macroscopic samples (such as industrial catalysts) which cannot feasibly be measured with lower energies due to high attenuation of the beam. For this, highly efficient focusing lens arrays will be essential, which will match the current development trend of nano-lithography optics -- high NA small aperture lenses. Another aspect here is the exploration of so-called "pink beam" measurements. Taking into account the capabilities of new sources in producing more temporally coherent beams, MBP may, with careful execution, increase the speed of data acquisition even further by utilizing currently wasted photons of different energy without requiring significant or perhaps even any monochromatization of the beam. 

Another direction of development for MBP is utilization at low-brilliance sources, such as older synchrotrons or bright laboratory sources. The latter are very attractive as they are widely available and, unlike user facilities, have easy access and affordable maintenance costs for a single scientific group. The knowledge and developments acquired at SRFs can in principle be almost directly applied to implement MBP for high-resolution quantitative 3D imaging in the laboratory, which opens up new avenues in many fields of science and industry and also significantly reduces the costs of measurements.

In summary, we demonstrate MBP as a high value and high performance method to image extended samples at the nanoscale. This was achieved due to spectacular progress in 3D lithography which allowed us to manufacture precise and highly efficient tailored optics with integrated coding phase plates.
In comparison to conventional hard X-ray ptychography, which currently offers the highest possible spatial resolution of known X-ray methods, MBP improves either field of view or scan speed with no compromize on spatial resolution. Due to the broad application fields of hard X-ray nano-imaging, and with the efficient preparation of suitable optics, we anticipate that MBP may in principle supersede the use of single beam ptychography for the study of large complex or composite samples.


\section*{Materials and Methods}

\subsection*{Lens array design and manufacturing}
\begin{figure}
    \centering
    \includegraphics[width=0.65\linewidth]{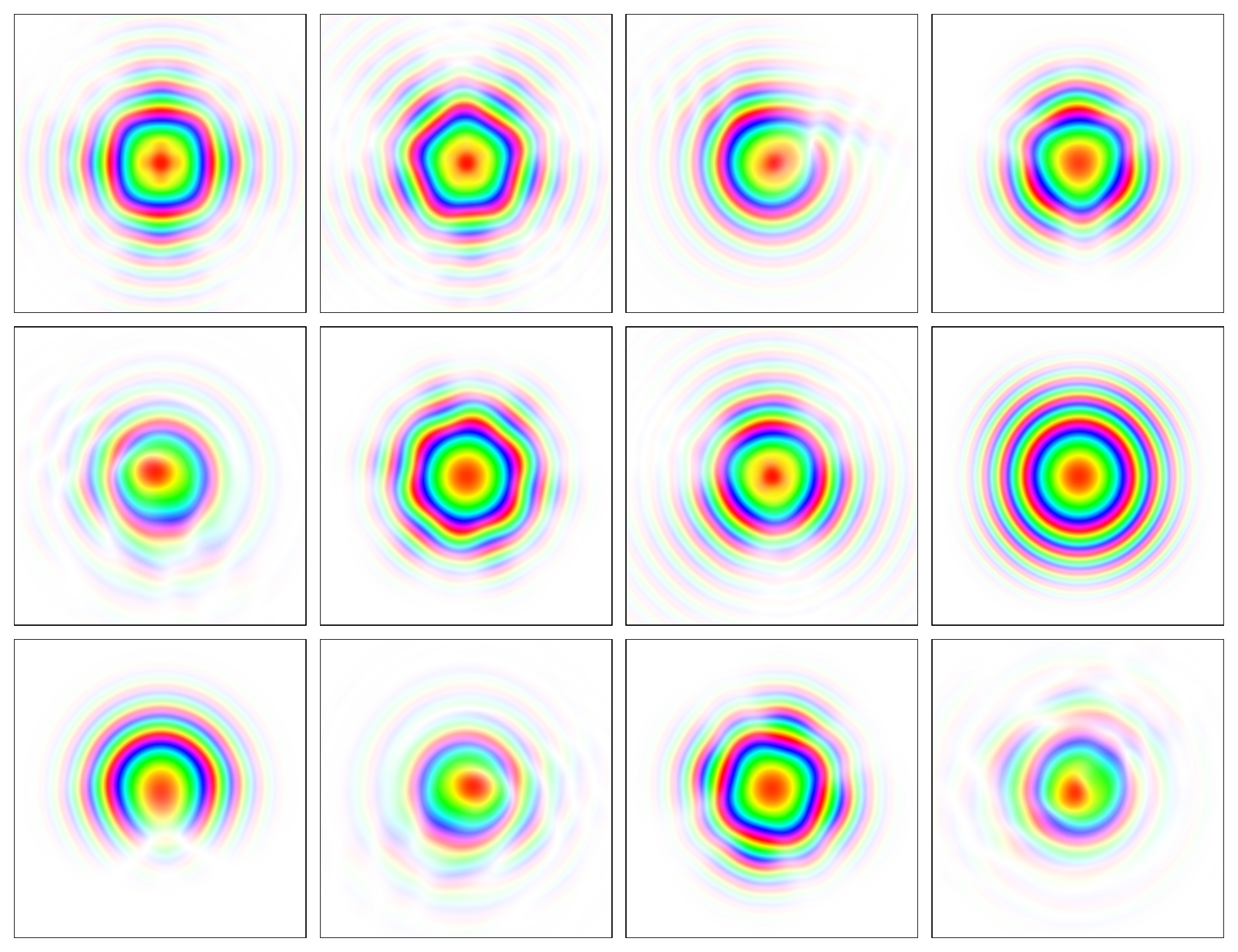}
    \caption{The simulated probe complex field cross-section with color-coded phase at \SI{5}{\milli\meter} defocus distance at \SI{13}{\kilo\eV}.}
    \label{fig:color-codedPP}
\end{figure}

For robust reconstruction, each beam needs to be uniquely coded in phase and/or amplitude to help in separation of superposed scattering signals at the detector. The criterion for phase plate design is that the amplitude on the defocus plane is tightly distributed, and the phase is distinguishable from probe to probe. Therefore, we designed 3 types of phase plate modal (the cake pieces, pyramid and vortex layers). Before manufacturing, we perform the simulation to check the probe difference based on the proper probe size on the sample plane. In Fig. \ref{fig:color-codedPP}, we show the probe amplitude with color-coded phase at \SI{5}{\milli\meter} defocus distance from the focal plane. The probe size is about \SI{1.5}{\micro\meter}. The height of phase plate designs based on the refractive index of IP-S materials and the maximum phase shift that the phase plate can bring is approximately 
$\pi$. For the design of the phase plate at \SI{13}{\kilo\eV}, the highest structure is about \SI{23}{\micro\meter}. For the design of phase plate at \SI{20}{\kilo\eV}, the highest structure is about \SI{45}{\micro\meter}.

Lens arrays and phase plates were fabricated using a commercial 3D laser printing setup (Nanoscribe, Photonics Professional GT) and a commercial photoresist system (Nanoscribe, IP-S) based on acrylates. We used a $25\times$ NA1.4 microscope objective lens that provides smooth surfaces when printing micro-optical components. The resulting lateral (axial) voxel size was approximately \SI{400}{\nano\meter} (\SI{2300}{\nano\meter}). We printed the structures directly on SiN membranes (Norcada NX10100D). Before printing, the membranes were cleaned in a plasma oven and subsequently silanized for better adhesion. All structures were printed with a slicing distance of \SI{300}{nm}, hatching distance of \SI{200}{\nano\meter}, focus-scanning speed of \SI{75}{\milli\meter/\sec}, and laser power of \SI{25}{\milli\W} (measured at the entrance pupil of the microscope objective lens).
Since the focal length of the central lens towers was different, the probe size at the sample plane was different as well. To correct for this in the second manufacturing step, the curvature of the two central lens towers was corrected to match the focal length of the others. A second lens array for 20 keV was manufactured in a similar fashion, but containing 40 lenses in a row instead of 30 to compensate for the higher incident beam energy. An important parameter in lens array manufacturing is the position of each individual lens with respect to the others over the whole height of the structure (see Fig. \ref{fig:sem}(a)). Since the structures have an aspect ratio exceeding 100:1, \SI{200}{\micro\radian} precision with respect to the X-ray beam is essential for proper alignment to maintain sufficient focusing performance. This constrains the position difference between the first and last lenses in each row with respect to the optical axis to \SI{0.6}{\micro\meter}, which was successfully achieved.
To ensure precision printing of the structure of the integrated phase plate, they were characterized during the manufacturing process using in-situ QPI technique \cite{qpi}.
It reconstructs the phase distribution of the structure from the wide-field intensity image stack. The images are acquired after printing but before development in the same volume of photoresist using LED with a wavelength of \SI{630}{\nano\meter}. The reconstructed phase distribution corresponds to the amount of printed material and is proportional to the product of the refractive index difference between unpolymerized and polymerized photoresist and the height of the printed structure.

\subsection*{Ptychographic experiments}
\subsubsection*{Beamline P06 PETRA III}
Partial results of this paper are taken from the nanoprobe end-station of beamline P06 of PETRA III at DESY in Hamburg, Germany. We used a 30-fold 3x4 lens tower array to create multiple probes by focusing X-rays, lenses were arranged in a grid with a spacing of \SI{30}{\micro\meter}. The sample was placed about \SI{70}{\milli\meter} downstream from the lens array. The detector-sample distance was \SI{308}{\centi\meter}. We use the Eiger (2048x2048 pix) as the detector with \SI{75}{\micro\meter} pixel size. The detector was placed in a vacuum to suppress the air scattering. The scanning was performed in a raster pattern with jittering of positions of \SI{20}{\percent} of the step size. The step size of \SI{350}{\nano\meter} was set according to the smallest probe size from the beams array at the sample position -- \SI{560}{\nano\meter}.  The scanning range of \SI{35}{\micro\meter} was chosen to ensure sufficient overlapping between areas scanned by adjacent probes. 
The dwell time for 12 beams experiment was chosen according to the detector saturation, namely 0.1s per scan point. For characterization of individual beams, we performed several measurements with lens tower combinations 2x2 and 1x1. The step size ranged from \SI{350}{\nano\meter} to \SI{750}{\nano\meter} and the dwell time ranged from \SI{0.25}{\s} to \SI{0.5}{\s} per scan point correspondingly. 

\subsubsection*{Beamline ID13 ESRF}

Other results of this paper is taken from the third end-station of ID13 beamline of the ESRF-EBS in Grenoble, France. We used the 40-fold 3x4 lens tower array with phase plates to create multiple probes by focusing X-rays at 20 keV with the design similar to the one used for 13 keV experiment. The sample was placed about \SI{108}{\milli\meter} downstream of the lens array. The detector-sample distance was \SI{5.03}{\meter}. The scan pattern was Fermat spiral mode\cite{Huang:14}. The position precision was about \SI{100}{\nano\meter}. We used the MAXIPIX (516x516 pix) as the detector with \SI{55}{\micro\meter} pixel size and placed it in the air. The detector was operated in single-exposure mode with a maximum photon count of 12518. The dwell time for for 6 beams experiment was chosen according to the detector saturation, namely 0.1s per scan point. For characterization of single beams the step size was \SI{500}{\nano\meter} and the dwell time was 0.2s per scan point. Therefore, our resolution and reconstruction is limited by the SNR and position accuracy.The scanning range of \SI{35}{\micro\meter} was chosen to ensure sufficient overlapping between areas scanned by neighboring probes. 

\subsection*{Ptychographic reconstructions}

\subsubsection*{Multibeam model}
For initial reconstructions we have adapted multiplex approach from Batey et. al\cite{Batey2014} to reconstruct spatially separated beams instead of different wavelengths. Here we briefly explain the difference between conventional single beam ptychography ePIE algorithm\cite{ePIE} and its multibeam counterpart. 
A single beam Ptychographic model describes the scattered wave-field at the detector plane as multiplication of the object and the probe at $j$ scan position of the sample plane:
\begin{equation}
\psi_{(x,y)} = O_{x,y} \times P_{(x-x_j,y-y_j)}
\end{equation}
whereas, for the multibeam case, $m$ probes with offsets $S_m$ illuminate the object:
\begin{equation}
\psi_{(x,y),m} = O_{x,y} \times P_{(x-x_j,y-y_j)-S_m}
\end{equation}
The propagated wavefront from each beam at the detector plane can be described in the form of Fourier transform:
\begin{equation}
\Psi = \mathcal{F}(\psi)
\end{equation}
Then, the intensity recorded at the detector in the case of a single beam described as:
\begin{equation}
I = |\Psi|^2,
\end{equation}
and, in case of multiple incoherent to each other beams, they add up incoherently:
\begin{equation}
I = \sum_m |\Psi_m|^2
\end{equation}
The update functions for object and probes in multibeam case:
\begin{equation}
O_{l+1,(x,y)} = O_{l,(x,y)} + \alpha\times\sum_m O_{l,(x,y)}\frac{P^*_{l,m,(x-x_j,y-y_j)-S_m}}{max|P_{l,(x-x_j,y-y_j)-S_m}|^2}\times-\phi_{j,m,(x,y)}
\end{equation}
\begin{equation}
P_{l+1,m,(x,y)}= P_{l,m,(x,y)} +\beta\times\frac{O^*_{l,(x+x_j,y+y_j)+S_m}}{max|O_{l,(x+x_j,y+y_j)+S_m}|^2}\times-\phi_{j,m,(x+x_j,y+y_j)+S_m},
\end{equation}
\begin{equation}
\phi_m = \mathcal{F}^{-1}((1 - \frac{\sqrt{n_j}}{\sqrt{I_j}})\Psi_m),
\end{equation}
here, for single beam $m = 1$ and $S_m = 0$,  $n_j$ is recorded diffraction intensity, $\alpha$ is the object update strength and $\beta$ is the illumination update strength.

It is worth noting that neighboring beams can be coherent to each other and in this case interference between them must be taken into account, or, it can be ignored if beams spacing $S_m$ set in the way that it satisfies alias cloaking condition:
\begin{equation}
S_m = nt\frac{d\lambda}{p},
\end{equation}
where $p$ is a detector pixel size, $\lambda$ is X-rays wavelength, $d$ is a sample-detector distance, and $nt$ is an integer.

\subsubsection*{Reconstruction approach}
For reconstructions we have used two software packages, internally build Ptycho and open-source PtyPy\cite{ptypy}. Before we perform multibeam reconstruction, we first chracterize each probe created by a single lens stack and retrieve the probe from single-beam ptychographic reconstruction with the standard ePIE algorithm\cite{ePIE} for 1,000 iterations. For the Siemens star reconstruction results from p06, the recorded far-field diffraction patters were cropped to 512 x 512 pixels centered around the beam axis. The pixel size in the reconstruction is \SI{8.1}{\nano\meter}. The image was reconstructed using 6,000 iterations of the ePIE algorithm with the object update strength $\alpha = 0.1$ and the illumination update strength $\beta = 1.0$, followed by 8,000 iterations with the stronger object update strength $\alpha = 1.0$ and the weaker illumination update strength $\beta = 0.1$ to further optimize object reconstruction. 

For the other reconstruction results from p06, the recorded far-field diffraction patters were cropped to 256 x 256 pixels centered around the beam axis. The pixel size in the reconstruction is \SI{16.2}{\nano\meter}. Every reconstruction consisted of 2500 iterations of the difference map (DM) algorithm\cite{DM_Algorithm} followed by 7500 iterations of the maximum-likelihood algorithm\cite{ML_Algorithm}.

For the reconstruction results from ID13, we used 4 probe modes for the 6 beams reconstruction to account for some fast beam oscillations. The recorded far-field diffraction patters were cropped to 128 x 128 pixels centered around the beam axis. The pixel size in the reconstruction is \SI{44.3}{\nano\meter}. Because we performed this experiment at a higher energy, the size of the probes illuminating the sample increased. We changed the reconstruction scheme by enlarging the probe field of view with a 2 x 2 upsampling \cite{upscaling_ptycho, vertualenlarge_ptycho}. Every reconstruction consisted of 1000 iterations of the difference map (DM) algorithmfollowed by 9000 iterations of the maximum-likelihood algorithm. 

In all reconstructions, to compensate for positioning errors in the scanning stage, the scan positions were numerically refined every 50 iterations to improve the reconstruction\cite{Maiden_positionrefine,ptycho_posrefine}.


\bibliography{bibliography}

\begin{thebibliography}{10}
\urlstyle{rm}
\expandafter\ifx\csname url\endcsname\relax
  \def\url#1{\texttt{#1}}\fi
\expandafter\ifx\csname urlprefix\endcsname\relax\def\urlprefix{URL }\fi
\expandafter\ifx\csname doiprefix\endcsname\relax\def\doiprefix{DOI: }\fi
\providecommand{\bibinfo}[2]{#2}
\providecommand{\eprint}[2][]{\url{#2}}

\bibitem{Cat_vis}
\bibinfo{author}{Werny, M.~J.}, \bibinfo{author}{Meirer, F.} \&
  \bibinfo{author}{Weckhuysen, B.~M.}
\newblock \bibinfo{journal}{\bibinfo{title}{Visualizing the structure,
  composition and activity of single catalyst particles for olefin
  polymerization and polyolefin decomposition}}.
\newblock {\emph{\JournalTitle{Angewandte Chemie International Edition}}}
  \textbf{\bibinfo{volume}{n/a}}, \bibinfo{pages}{e202306033},
  \doiprefix\url{https://doi.org/10.1002/anie.202306033}
  (\bibinfo{year}{2023}).
\newblock
  \eprint{https://onlinelibrary.wiley.com/doi/pdf/10.1002/anie.202306033}.

\bibitem{das_review}
\bibinfo{author}{Das, S.}, \bibinfo{author}{Pashminehazar, R.},
  \bibinfo{author}{Sharma, S.}, \bibinfo{author}{Weber, S.} \&
  \bibinfo{author}{Sheppard, T.~L.}
\newblock \bibinfo{journal}{\bibinfo{title}{New dimensions in catalysis
  research with hard x-ray tomography}}.
\newblock {\emph{\JournalTitle{Chemie Ingenieur Technik}}}
  \textbf{\bibinfo{volume}{94}}, \bibinfo{pages}{1591--1610},
  \doiprefix\url{https://doi.org/10.1002/cite.202200082}
  (\bibinfo{year}{2022}).
\newblock
  \eprint{https://onlinelibrary.wiley.com/doi/pdf/10.1002/cite.202200082}.

\bibitem{zenyuk_review}
\bibinfo{author}{Chen, Y.} \emph{et~al.}
\newblock \bibinfo{journal}{\bibinfo{title}{A viewpoint on x-ray tomography
  imaging in electrocatalysis}}.
\newblock {\emph{\JournalTitle{ACS Catalysis}}} \textbf{\bibinfo{volume}{13}},
  \bibinfo{pages}{10010--10025}, \doiprefix\url{10.1021/acscatal.3c01453}
  (\bibinfo{year}{2023}).
\newblock \eprint{https://doi.org/10.1021/acscatal.3c01453}.

\bibitem{Miao1999}
\bibinfo{author}{Miao, J.}, \bibinfo{author}{Charalambous, P.},
  \bibinfo{author}{Kirz, J.} \& \bibinfo{author}{Sayre, D.}
\newblock \bibinfo{journal}{\bibinfo{title}{Extending the methodology of x-ray
  crystallography to allow imaging of micrometre-sized non-crystalline
  specimens}}.
\newblock {\emph{\JournalTitle{Nature}}} \textbf{\bibinfo{volume}{400}},
  \bibinfo{pages}{342--344}, \doiprefix\url{10.1038/22498}
  (\bibinfo{year}{1999}).

\bibitem{Chapman2010}
\bibinfo{author}{Chapman, H.~N.} \& \bibinfo{author}{Nugent, K.~A.}
\newblock \bibinfo{journal}{\bibinfo{title}{Coherent lensless x-ray imaging}}.
\newblock {\emph{\JournalTitle{Nature Photonics}}}
  \textbf{\bibinfo{volume}{4}}, \bibinfo{pages}{833--839},
  \doiprefix\url{10.1038/nphoton.2010.240} (\bibinfo{year}{2010}).

\bibitem{Dierolf2010}
\bibinfo{author}{Dierolf, M.} \emph{et~al.}
\newblock \bibinfo{journal}{\bibinfo{title}{Ptychographic x-ray computed
  tomography at the nanoscale}}.
\newblock {\emph{\JournalTitle{Nature}}} \textbf{\bibinfo{volume}{467}},
  \bibinfo{pages}{436--439}, \doiprefix\url{10.1038/nature09419}
  (\bibinfo{year}{2010}).

\bibitem{Pfeiffer2018}
\bibinfo{author}{Pfeiffer, F.}
\newblock \bibinfo{journal}{\bibinfo{title}{X-ray ptychography}}.
\newblock {\emph{\JournalTitle{Nature Photonics}}}
  \textbf{\bibinfo{volume}{12}}, \bibinfo{pages}{9--17},
  \doiprefix\url{10.1038/s41566-017-0072-5} (\bibinfo{year}{2018}).

\bibitem{grote2022}
\bibinfo{author}{Grote, L.} \emph{et~al.}
\newblock \bibinfo{journal}{\bibinfo{title}{Imaging cu2o nanocube hollowing in
  solution by quantitative in situ x-ray ptychography}}.
\newblock {\emph{\JournalTitle{Nature Communications}}}
  \textbf{\bibinfo{volume}{13}}, \bibinfo{pages}{4971},
  \doiprefix\url{https://doi.org/10.1038/s41467-022-32373-2}
  (\bibinfo{year}{2022}).

\bibitem{Taphorn2022}
\bibinfo{author}{Taphorn, K.} \emph{et~al.}
\newblock \bibinfo{journal}{\bibinfo{title}{X-ray stain localization with
  near-field ptychographic computed tomography}}.
\newblock {\emph{\JournalTitle{Advanced Science}}}
  \textbf{\bibinfo{volume}{9}}, \bibinfo{pages}{2201723},
  \doiprefix\url{https://doi.org/10.1002/advs.202201723}
  (\bibinfo{year}{2022}).
\newblock
  \eprint{https://onlinelibrary.wiley.com/doi/pdf/10.1002/advs.202201723}.

\bibitem{Cryo_electron_ptycho2023}
\bibinfo{author}{Pei, X.} \emph{et~al.}
\newblock \bibinfo{journal}{\bibinfo{title}{Cryogenic electron ptychographic
  single particle analysis with wide bandwidth information transfer}}.
\newblock {\emph{\JournalTitle{Nature Communications}}}
  \textbf{\bibinfo{volume}{14}}, \bibinfo{pages}{3027},
  \doiprefix\url{10.1038/s41467-023-38268-0} (\bibinfo{year}{2023}).

\bibitem{Ding2022}
\bibinfo{author}{Ding, Z.} \emph{et~al.}
\newblock \bibinfo{journal}{\bibinfo{title}{Three-dimensional electron
  ptychography of organic--inorganic hybrid nanostructures}}.
\newblock {\emph{\JournalTitle{Nature Communications}}}
  \textbf{\bibinfo{volume}{13}}, \bibinfo{pages}{4787},
  \doiprefix\url{10.1038/s41467-022-32548-x} (\bibinfo{year}{2022}).

\bibitem{daSilva:gb5085}
\bibinfo{author}{da~Silva, J.~C.} \emph{et~al.}
\newblock \bibinfo{journal}{\bibinfo{title}{{Overcoming the challenges of
  high-energy X-ray ptychography}}}.
\newblock {\emph{\JournalTitle{Journal of Synchrotron Radiation}}}
  \textbf{\bibinfo{volume}{26}}, \bibinfo{pages}{1751--1762},
  \doiprefix\url{10.1107/S1600577519006301} (\bibinfo{year}{2019}).

\bibitem{bevis2017}
\bibinfo{author}{Bevis, C.} \emph{et~al.}
\newblock \bibinfo{journal}{\bibinfo{title}{Multiple beam ptychography for
  large field-of-view, high throughput, quantitative phase contrast imaging}}.
\newblock {\emph{\JournalTitle{Ultramicroscopy}}} \textbf{\bibinfo{volume}{184,
  Part A}}, \bibinfo{pages}{164 -- 171},
  \doiprefix\url{10.1016/j.ultramic.2017.08.018} (\bibinfo{year}{2017}).

\bibitem{Yao2020}
\bibinfo{author}{Yao, Y.} \emph{et~al.}
\newblock \bibinfo{journal}{\bibinfo{title}{Multi-beam {{X}}-ray ptychography
  for high-throughput coherent diffraction imaging}}.
\newblock {\emph{\JournalTitle{Scientific Reports}}}
  \textbf{\bibinfo{volume}{10}}, \doiprefix\url{10.1038/s41598-020-76412-8}
  (\bibinfo{year}{2020}).

\bibitem{Hirose2020}
\bibinfo{author}{Hirose, M.}, \bibinfo{author}{Higashino, T.},
  \bibinfo{author}{Ishiguro, N.} \& \bibinfo{author}{Takahashi, Y.}
\newblock \bibinfo{journal}{\bibinfo{title}{Multibeam ptychography with
  synchrotron hard {{X}}-rays}}.
\newblock {\emph{\JournalTitle{Optics Express}}} \textbf{\bibinfo{volume}{28}},
  \bibinfo{pages}{1216}, \doiprefix\url{10.1364/OE.378083}
  (\bibinfo{year}{2020}).

\bibitem{Wittwer21}
\bibinfo{author}{Wittwer, F.} \emph{et~al.}
\newblock \bibinfo{journal}{\bibinfo{title}{{Upscaling of multi-beam x-ray
  ptychography for efficient x-ray microscopy with high resolution and large
  field of view}}}.
\newblock {\emph{\JournalTitle{Applied Physics Letters}}}
  \textbf{\bibinfo{volume}{118}}, \bibinfo{pages}{171102},
  \doiprefix\url{10.1063/5.0045571} (\bibinfo{year}{2021}).
\newblock
  \eprint{https://pubs.aip.org/aip/apl/article-pdf/doi/10.1063/5.0045571/14547517/171102\_1\_online.pdf}.

\bibitem{Lyubomirskiy22}
\bibinfo{author}{Lyubomirskiy, M.} \emph{et~al.}
\newblock \bibinfo{journal}{\bibinfo{title}{Multi-beam x-ray ptychography using
  coded probes for rapid non-destructive high resolution imaging of extended
  samples}}.
\newblock {\emph{\JournalTitle{Sci. Rep.}}} \textbf{\bibinfo{volume}{26}},
  \bibinfo{pages}{6203},
  \doiprefix\url{https://doi.org/10.1038/s41598-022-09466-5}
  (\bibinfo{year}{2022}).

\bibitem{Gorelick2011-rh}
\bibinfo{author}{Gorelick, S.} \emph{et~al.}
\newblock \bibinfo{journal}{\bibinfo{title}{High-efficiency fresnel zone plates
  for hard x-rays by 100 kev e-beam lithography and electroplating}}.
\newblock {\emph{\JournalTitle{J Synchrotron Radiat}}}
  \textbf{\bibinfo{volume}{18}}, \bibinfo{pages}{442--446},
  \doiprefix\url{10.1107/S0909049511002366} (\bibinfo{year}{2011}).

\bibitem{Lengeler2002}
\bibinfo{author}{Lengeler, B.} \emph{et~al.}
\newblock \bibinfo{journal}{\bibinfo{title}{{Parabolic refractive X-ray
  lenses}}}.
\newblock {\emph{\JournalTitle{Journal of Synchrotron Radiation}}}
  \textbf{\bibinfo{volume}{9}}, \bibinfo{pages}{119--124},
  \doiprefix\url{10.1107/S0909049502003436} (\bibinfo{year}{2002}).

\bibitem{Schroer2003}
\bibinfo{author}{Schroer, C.~G.} \emph{et~al.}
\newblock \bibinfo{journal}{\bibinfo{title}{{Nanofocusing parabolic refractive
  x-ray lenses}}}.
\newblock {\emph{\JournalTitle{Applied Physics Letters}}}
  \textbf{\bibinfo{volume}{82}}, \bibinfo{pages}{1485--1487},
  \doiprefix\url{10.1063/1.1556960} (\bibinfo{year}{2003}).
\newblock
  \eprint{https://pubs.aip.org/aip/apl/article-pdf/82/9/1485/7827613/1485\_1\_online.pdf}.

\bibitem{Petrov17}
\bibinfo{author}{Petrov, A.~K.} \emph{et~al.}
\newblock \bibinfo{journal}{\bibinfo{title}{Polymer x-ray refractive
  nano-lenses fabricated by additive technology}}.
\newblock {\emph{\JournalTitle{Opt. Express}}} \textbf{\bibinfo{volume}{25}},
  \bibinfo{pages}{14173--14181}, \doiprefix\url{10.1364/OE.25.014173}
  (\bibinfo{year}{2017}).

\bibitem{Lyubomirskiy2019}
\bibinfo{author}{Lyubomirskiy, M.} \emph{et~al.}
\newblock \bibinfo{journal}{\bibinfo{title}{Ptychographic characterisation of
  polymer compound refractive lenses manufactured by additive technology}}.
\newblock {\emph{\JournalTitle{Optics Express}}} \textbf{\bibinfo{volume}{27}},
  \bibinfo{pages}{8639}, \doiprefix\url{10.1364/OE.27.008639}
  (\bibinfo{year}{2019}).

\bibitem{Seiboth19}
\bibinfo{author}{Seiboth, F.} \emph{et~al.}
\newblock \bibinfo{journal}{\bibinfo{title}{Refractive hard x-ray vortex phase
  plates}}.
\newblock {\emph{\JournalTitle{Opt. Lett.}}} \textbf{\bibinfo{volume}{44}},
  \bibinfo{pages}{4622--4625}, \doiprefix\url{10.1364/OL.44.004622}
  (\bibinfo{year}{2019}).

\bibitem{Wegener22}
\bibinfo{author}{Wegener, M.}
\newblock \bibinfo{journal}{\bibinfo{title}{3d laser nanoprinting: Recent
  progress}}.
\newblock {\emph{\JournalTitle{Proceedings of the 2022 Conference on Lasers and
  Electro-Optics Pacific Rim}}} \doiprefix\url{10.1364/CLEOPR.2022.CMP16B_01}
  (\bibinfo{year}{2022}).

\bibitem{qpi}
\bibinfo{author}{Zvagelsky, R.}, \bibinfo{author}{Kiefer, P.},
  \bibinfo{author}{Weinacker, J.} \& \bibinfo{author}{Wegener, M.}
\newblock \bibinfo{journal}{\bibinfo{title}{In-situ quantitative phase imaging
  during multi-photon laser printing}}.
\newblock {\emph{\JournalTitle{ACS Photonics}}} \textbf{\bibinfo{volume}{10}},
  \bibinfo{pages}{2901--2908}, \doiprefix\url{10.1021/acsphotonics.3c00625}
  (\bibinfo{year}{2023}).

\bibitem{XRESO}
\bibinfo{author}{NTT-AT}.
\newblock \bibinfo{title}{Manufacturer page}.
\newblock \bibinfo{howpublished}{\url{www.ntt-at.com/product/x-ray_chart/}}
  (\bibinfo{year}{2023}).

\bibitem{weber_tomo}
\bibinfo{author}{Weber, S.} \emph{et~al.}
\newblock \bibinfo{journal}{\bibinfo{title}{Evolution of hierarchically porous
  nickel alumina catalysts studied by x-ray ptychography}}.
\newblock {\emph{\JournalTitle{Advanced Science}}}
  \textbf{\bibinfo{volume}{9}}, \bibinfo{pages}{2105432},
  \doiprefix\url{https://doi.org/10.1002/advs.202105432}
  (\bibinfo{year}{2022}).
\newblock
  \eprint{https://onlinelibrary.wiley.com/doi/pdf/10.1002/advs.202105432}.

\bibitem{CITIUS}
\bibinfo{author}{Takahashi, Y.} \emph{et~al.}
\newblock \bibinfo{journal}{\bibinfo{title}{{High-resolution and
  high-sensitivity X-ray ptychographic coherent diffraction imaging using the
  CITIUS detector}}}.
\newblock {\emph{\JournalTitle{Journal of Synchrotron Radiation}}}
  \textbf{\bibinfo{volume}{30}}, \bibinfo{pages}{989--994},
  \doiprefix\url{10.1107/S1600577523004897} (\bibinfo{year}{2023}).

\bibitem{JUNGFRAU}
\bibinfo{author}{Leonarski, F.} \emph{et~al.}
\newblock \bibinfo{journal}{\bibinfo{title}{{Jungfraujoch: hardware-accelerated
  data-acquisition system for kilohertz pixel-array X-ray detectors}}}.
\newblock {\emph{\JournalTitle{Journal of Synchrotron Radiation}}}
  \textbf{\bibinfo{volume}{30}}, \bibinfo{pages}{227--234},
  \doiprefix\url{10.1107/S1600577522010268} (\bibinfo{year}{2023}).

\bibitem{batey2022}
\bibinfo{author}{Batey, D.}, \bibinfo{author}{Rau, C.} \&
  \bibinfo{author}{Cipiccia, S.}
\newblock \bibinfo{journal}{\bibinfo{title}{High-speed x-ray ptychographic
  tomography}}.
\newblock {\emph{\JournalTitle{Scientific Reports}}}
  \textbf{\bibinfo{volume}{12}}, \bibinfo{pages}{7846} (\bibinfo{year}{2022}).

\bibitem{meirer2018}
\bibinfo{author}{Meirer, F.} \& \bibinfo{author}{Weckhuysen, B.~M.}
\newblock \bibinfo{journal}{\bibinfo{title}{Spatial and temporal exploration of
  heterogeneous catalysts with synchrotron radiation}}.
\newblock {\emph{\JournalTitle{Nature Reviews Materials}}}
  \textbf{\bibinfo{volume}{3}}, \bibinfo{pages}{324--340}
  (\bibinfo{year}{2018}).

\bibitem{Bossers2021}
\bibinfo{author}{Bossers, K.~W.} \emph{et~al.}
\newblock \bibinfo{journal}{\bibinfo{title}{Heterogeneity in the fragmentation
  of ziegler catalyst particles during ethylene polymerization quantified by
  x-ray nanotomography}}.
\newblock {\emph{\JournalTitle{JACS Au}}} \textbf{\bibinfo{volume}{1}},
  \bibinfo{pages}{852--864}, \doiprefix\url{10.1021/jacsau.1c00130}
  (\bibinfo{year}{2021}).
\newblock \bibinfo{note}{PMID: 34240080},
  \eprint{https://doi.org/10.1021/jacsau.1c00130}.

\bibitem{Bare2014}
\bibinfo{author}{Bare, S.~R.} \emph{et~al.}
\newblock \bibinfo{journal}{\bibinfo{title}{Characterization of a fluidized
  catalytic cracking catalyst on ensemble and individual particle level by
  x-ray micro- and nanotomography, micro-x-ray fluorescence, and micro-x-ray
  diffraction}}.
\newblock {\emph{\JournalTitle{ChemCatChem}}} \textbf{\bibinfo{volume}{6}},
  \bibinfo{pages}{1427--1437},
  \doiprefix\url{https://doi.org/10.1002/cctc.201300974}
  (\bibinfo{year}{2014}).
\newblock
  \eprint{https://chemistry-europe.onlinelibrary.wiley.com/doi/pdf/10.1002/cctc.201300974}.

\bibitem{MAXIV}
\bibinfo{author}{Robert, A.} \emph{et~al.}
\newblock \bibinfo{journal}{\bibinfo{title}{Max iv laboratory}}.
\newblock {\emph{\JournalTitle{The European Physical Journal Plus}}}
  \textbf{\bibinfo{volume}{138}}, \bibinfo{pages}{495},
  \doiprefix\url{10.1140/epjp/s13360-023-04018-w} (\bibinfo{year}{2023}).

\bibitem{PETRAIV}
\bibinfo{author}{Schroer, C.~G.} \emph{et~al.}
\newblock \bibinfo{journal}{\bibinfo{title}{The synchrotron radiation source
  petra iii and its future ultra-low-emittance upgrade petra iv}}.
\newblock {\emph{\JournalTitle{The European Physical Journal Plus}}}
  \textbf{\bibinfo{volume}{137}}, \bibinfo{pages}{1312},
  \doiprefix\url{10.1140/epjp/s13360-022-03517-6} (\bibinfo{year}{2022}).

\bibitem{Huang:14}
\bibinfo{author}{Huang, X.} \emph{et~al.}
\newblock \bibinfo{journal}{\bibinfo{title}{Optimization of overlap uniformness
  for ptychography}}.
\newblock {\emph{\JournalTitle{Opt. Express}}} \textbf{\bibinfo{volume}{22}},
  \bibinfo{pages}{12634--12644}, \doiprefix\url{10.1364/OE.22.012634}
  (\bibinfo{year}{2014}).

\bibitem{Batey2014}
\bibinfo{author}{Batey, D.~J.}, \bibinfo{author}{Claus, D.} \&
  \bibinfo{author}{Rodenburg, J.~M.}
\newblock \bibinfo{journal}{\bibinfo{title}{Information multiplexing in
  ptychography}}.
\newblock {\emph{\JournalTitle{Ultramicroscopy}}}
  \textbf{\bibinfo{volume}{138}}, \bibinfo{pages}{13--21},
  \doiprefix\url{https://doi.org/10.1016/j.ultramic.2013.12.003}
  (\bibinfo{year}{2014}).

\bibitem{ePIE}
\bibinfo{author}{Maiden, A.~M.} \& \bibinfo{author}{Rodenburg, J.~M.}
\newblock \bibinfo{journal}{\bibinfo{title}{An improved ptychographical phase
  retrieval algorithm for diffractive imaging}}.
\newblock {\emph{\JournalTitle{Ultramicroscopy}}}
  \textbf{\bibinfo{volume}{109}}, \bibinfo{pages}{1256--1262},
  \doiprefix\url{https://doi.org/10.1016/j.ultramic.2009.05.012}
  (\bibinfo{year}{2009}).

\bibitem{ptypy}
\bibinfo{author}{Enders, B.} \& \bibinfo{author}{Thibault, P.}
\newblock \bibinfo{journal}{\bibinfo{title}{A computational framework for
  ptychographic reconstructions}}.
\newblock {\emph{\JournalTitle{Proceedings of the Royal Society A:
  Mathematical, Physical and Engineering Sciences}}}
  \textbf{\bibinfo{volume}{472}}, \bibinfo{pages}{20160640},
  \doiprefix\url{10.1098/rspa.2016.0640} (\bibinfo{year}{2016}).
\newblock
  \eprint{https://royalsocietypublishing.org/doi/pdf/10.1098/rspa.2016.0640}.

\bibitem{DM_Algorithm}
\bibinfo{author}{Thibault, P.}, \bibinfo{author}{Dierolf, M.},
  \bibinfo{author}{Bunk, O.}, \bibinfo{author}{Menzel, A.} \&
  \bibinfo{author}{Pfeiffer, F.}
\newblock \bibinfo{journal}{\bibinfo{title}{Probe retrieval in ptychographic
  coherent diffractive imaging}}.
\newblock {\emph{\JournalTitle{Ultramicroscopy}}}
  \textbf{\bibinfo{volume}{109}}, \bibinfo{pages}{338--343},
  \doiprefix\url{https://doi.org/10.1016/j.ultramic.2008.12.011}
  (\bibinfo{year}{2009}).

\bibitem{ML_Algorithm}
\bibinfo{author}{Thibault, P.} \& \bibinfo{author}{Guizar-Sicairos, M.}
\newblock \bibinfo{journal}{\bibinfo{title}{Maximum-likelihood refinement for
  coherent diffractive imaging}}.
\newblock {\emph{\JournalTitle{New Journal of Physics}}}
  \textbf{\bibinfo{volume}{14}}, \bibinfo{pages}{063004},
  \doiprefix\url{10.1088/1367-2630/14/6/063004} (\bibinfo{year}{2012}).

\bibitem{upscaling_ptycho}
\bibinfo{author}{Batey, D.~J.} \emph{et~al.}
\newblock \bibinfo{journal}{\bibinfo{title}{Reciprocal-space up-sampling from
  real-space oversampling in x-ray ptychography}}.
\newblock {\emph{\JournalTitle{Phys. Rev. A}}} \textbf{\bibinfo{volume}{89}},
  \bibinfo{pages}{043812}, \doiprefix\url{10.1103/PhysRevA.89.043812}
  (\bibinfo{year}{2014}).

\bibitem{vertualenlarge_ptycho}
\bibinfo{author}{Wittwer, F.} \emph{et~al.}
\newblock \bibinfo{journal}{\bibinfo{title}{{Ptychography with a Virtually
  Enlarged Illumination.}}}
\newblock {\emph{\JournalTitle{Microscopy and Microanalysis}}}
  \textbf{\bibinfo{volume}{24}}, \bibinfo{pages}{46--47},
  \doiprefix\url{10.1017/S1431927618012667} (\bibinfo{year}{2018}).
\newblock
  \eprint{https://academic.oup.com/mam/article-pdf/24/S2/46/48093261/mam0046.pdf}.

\bibitem{Maiden_positionrefine}
\bibinfo{author}{Maiden, A.}, \bibinfo{author}{Humphry, M.},
  \bibinfo{author}{Sarahan, M.}, \bibinfo{author}{Kraus, B.} \&
  \bibinfo{author}{Rodenburg, J.}
\newblock \bibinfo{journal}{\bibinfo{title}{An annealing algorithm to correct
  positioning errors in ptychography}}.
\newblock {\emph{\JournalTitle{Ultramicroscopy}}}
  \textbf{\bibinfo{volume}{120}}, \bibinfo{pages}{64--72},
  \doiprefix\url{https://doi.org/10.1016/j.ultramic.2012.06.001}
  (\bibinfo{year}{2012}).

\bibitem{ptycho_posrefine}
\bibinfo{author}{Schropp, A.} \emph{et~al.}
\newblock \bibinfo{journal}{\bibinfo{title}{Full spatial characterization of a
  nanofocused x-ray free-electron laser beam by ptychographic imaging}}.
\newblock {\emph{\JournalTitle{Scientific reports}}}
  \textbf{\bibinfo{volume}{3}}, \bibinfo{pages}{1633},
  \doiprefix\url{https://doi.org/10.1038/srep01633} (\bibinfo{year}{2013}).

\end{thebibliography}

\section*{Acknowledgements}
The work was supported by Roentgen Angstrom Cluster grant (VR  2021-05975 and BMBF 13K22CHC). T.L.S. acknowledges funding from the Deutsche Forschungsgemeinschaft (DFG, German Research Foundation) in the framework of the Collaborative Research Centre SFB 1441 (project number 426888090, project C3). R. Z. and M. W. acknowledge funding from the Deutsche Forschungsgemeinschaft (DFG, German Research Foundation) under Germany’s Excellence Strategy 2082/1-390761711 (Excellence Cluster “3D Matter Made to Order”) and from the Karlsruhe School of Optics and Photonics (KSOP).  The authors are grateful to Thomas Keller and Satishkumar Kulkarni from DESY Nanolab for catalyst FIB preparation and Frank Seiboth for custom pinhole array manufacturing.

\section*{Author contributions statement}
M.L., T.L. conceived the experiments,  M.L., T.L., M.K., P.V.-P., R.Z., K.V.F., R.Y. conducted the experiments, T.L., M.K., M.L., P.V.-P. analysed the results, M.L., P.V.-P., M.K., T.L.S., M.W. acquired funding, all authors reviewed the manuscript. 

\end{document}